\documentclass{PoS}
\usepackage{amsmath}

\def\MeV{{\rm Me\!V}}

\def\gtwid{{\,\raise.3ex\hbox{$>$\kern-.75em\lower1ex\hbox{$\sim$}}\,}}
\def\ltwid{{\,\raise.3ex\hbox{$<$\kern-.75em\lower1ex\hbox{$\sim$}}\,}}

\def\chpt{\raise0.4ex\hbox{$\chi$}PT}
\def\schpt{S\raise0.4ex\hbox{$\chi$}PT}
\def\rschpt{rS\raise0.4ex\hbox{$\chi$}PT}

\title{Results for light pseudoscalar mesons}

\ShortTitle{Results for light pseudoscalar mesons}

\author{A.\ Bazavov, W.\ Freeman and D.\ Toussaint\\
        Department of Physics, University of Arizona, Tucson, AZ 85721, USA\\
	\phantom{email}}

\author{\speaker{C.\ Bernard} and X.\ Du \\
        Department of Physics, Washington University, St.~Louis, MO 63130, USA\\
        E-mail: \email{cb@wustl.edu}\\ }

\author{C.\ DeTar, L.\ Levkova and M.B.\ Oktay\\
        Physics Department, University of Utah, Salt Lake City, UT 84112, USA\\
	\phantom{email}}

\author{Steven Gottlieb\\
        Department of Physics, Indiana University, Bloomington, IN 47405, USA\\
        and NCSA, University of Illinois, Urbana, IL 61801, USA \\
	\phantom{email}}

\author{Urs M.\ Heller\\
        American Physical Society, One Research Road, Ridge, NY 11961, USA\\
	\phantom{email}}

\author{J.E.\ Hetrick\\
        Physics Department, University of the Pacific, Stockton, CA 95211, USA\\
	\phantom{email}}

\author{J.\ Laiho\\
	SUPA, School of Physics \& Astronomy, University of Glasgow, Glasgow G12 8QQ, UK\\
	\phantom{email}}

\author{J.\ Osborn\\
        Argonne Leadership Computing Facility, Argonne National Laboratory, Argonne, IL 60439, USA\\
	\phantom{email}}

\author{R.\ Sugar\\
        Department of Physics, University of California, Santa  Barbara, CA 93106, USA\\
	\phantom{email}}

\author{R.S.\ Van de Water\\
        Department of Physics, Brookhaven National Laboratory, Upton, NY 11973, USA\\
	\phantom{email}}

\vspace{-16mm}

\abstract{We present the current status
of the MILC collaboration's calculations of the properties of the light pseudoscalar
meson sector. We use asqtad staggered ensembles with 2+1 dynamical flavors
down to $a\!\approx\!0.045$ fm
and light quark mass down to $0.05m_s$.
Here we describe fits to the data using chiral forms from SU(3) chiral perturbation theory, including  all
staggered taste violations at NLO and the
continuum NNLO chiral logarithms. 
We emphasize issues of convergence of the chiral expansion.}

\FullConference{The XXVIII International Symposium on Lattice Field Theory, Lattice2010\\
		June 14-19, 2010\\
		Villasimius, Italy}

\begin{document}

\section{Introduction}

The MILC collaboration has been carrying out simulations of 2+1 flavor
lattice QCD with the improved ``asqtad'' staggered quark action since 1999;
for a review of the physics program 
see Ref.~\cite{Bazavov:2009bb}.  The asqtad ensembles are now complete,
and we are progressing to lattice generation with
 the more highly improved HISQ action \cite{HISQ},
with promising results so far \cite{HISQ-scaling}.
Extraction of physics with the asqtad ensembles is however continuing.
In particular, we are in the process of finishing
our (asqtad) study of the light pseudoscalar meson sector. Here we
give the latest update of this project, focusing on SU(3) chiral fits. 

Compared to the last status
reports in Refs.~\cite{SU3:2009,Xining:lat09}, we have completed
the final four ensembles, adding between 35 and 100\%
more configurations. The ensembles include two of the ones
with lighter-than-physical strange quark masses, which are crucial for SU(3) chiral fits.
This new data has allowed us to remove
all {\it a priori}\/ restrictions on SU(3) NNLO low energy
constants (LECs) from the fits. It is therefore now possible, for the
first time, to report physical values for some of these NNLO LECs, although the errors in
the values are
quite large. In addition,  we now consider additional alternative versions of the chiral fits,
including different treatments of still higher order effects.
These changes in the analysis give somewhat larger
systematic errors on the LECs at NLO, as well as on
the decay constant and quark condensate in the 
three-flavor chiral limit.  On the other hand,
errors on $f_\pi$ and $f_K$, and on quark masses, are largely insensitive to the
changes, and indeed have continued to decrease as the ensembles have improved.

We have also examined in detail the convergence of the SU(3) chiral expansion,
as a function of quark mass.  Results are presented both for the case
of three degenerate quark masses and for the more physical case of one quark (strange) much heavier
than the other two. The expansion appears
to be significantly better behaved in the former case than in the latter one.

\section{The ensembles and the fitting procedures}

The present analysis uses MILC ensembles 
at $a \approx 0.09$ fm,
$a \approx 0.06$ fm and $a \approx 0.045$ fm.
Although the MILC collaboration has also generated ensembles at three coarser
lattice spacings from $a \approx 0.18$ fm to $a \approx 0.12$ fm,
using that data in the fits is not currently feasible, since it would require analytic control
of higher order discretization effects. 
The ensembles used in this study are listed in
Table~\ref{tab:ensembles}. The quantity
$\hat m'$ is the simulated light ($u,d$) quark mass.
Three ensembles have an unphysically light simulated strange quark mass
$m'_s\approx0.6m_s$ (two $a\approx\!0.09$ ensembles with $am'_s=0.0186$,
and one $a\approx\!0.06$ ensemble with $am'_s=0.0108$), and one ensemble
has three degenerate quarks with $\hat m'=m_s'\approx0.1m_s$ (the
$a\approx\!0.09$ ensemble with  $a\hat m'=am'_s=0.0031$).
These light-$m'_s$ ensembles were created
specifically to help control the SU(3) \chpt\ fits.
\begin{table}
\begin{center}
\setlength{\tabcolsep}{1.0mm}
\begin{tabular}{|c|c|c|c|c|c|c|c|}
\hline
$a$ (fm) &$a\hat m'$ / $am'_s$ & $10/g^2$ & size & \# lats.& 
 $u_0$ & $r_1/a$ & $m_\pi L$ \\
\noalign{\vspace{-0.06cm}}
\hline
$\approx\!0.09$ & 0.0124 / 0.031  & 7.11 & $28^3\times96$ & 531 & 
 0.8788 & 3.858 & 5.78 \\
$\approx\!0.09$ & 0.0093 / 0.031  & 7.10 & $28^3\times96$ & 1124 & 
 0.8785 & 3.823 & 5.04 \\
$\approx\!0.09$ & 0.0062 / 0.031  & 7.09 & $28^3\times96$ & 591 & 
 0.8782 & 3.789 & 4.14 \\
$\approx\!0.09$ & 0.00465 / 0.031 & 7.085 & $32^3\times96$ & 984$^*$ & 
 0.8781 & 3.772 & 4.11 \\
$\approx\!0.09$ & 0.0031 / 0.031  & 7.08 & $40^3\times96$ & 945 & 
 0.8779 & 3.755 & 4.21 \\
$\approx\!0.09$ & 0.00155 / 0.031 & 7.075 & $64^3\times96$ & 751$^*$ & 
 0.877805 & 3.738 & 4.80 \\
\noalign{\vspace{-0.06cm}}
\hline
$\approx\!0.09$ & 0.0062  / 0.0186 & 7.10 & $28^3\times96$ & 985 & 
 0.8785 & 3.8823 & 4.09 \\
$\approx\!0.09$ & 0.0031  / 0.0186 & 7.06 & $40^3\times96$ & 781$^*$ &  
  0.8774 & 3.687 & 4.22 \\
$\approx\!0.09$ & 0.0031  / 0.0031 & 7.045 & $40^3\times96$ & 555$^*$ & 
 0.8770 & 3.637 & 4.20 \\
\noalign{\vspace{-0.06cm}}
\hline
$\approx\!0.06$ & 0.0072  / 0.018 & 7.48 & $48^3\times144$ & 594 & 
 0.8881 & 5.399 & 6.33 \\
$\approx\!0.06$ & 0.0054  / 0.018 & 7.475 & $48^3\times144$ & 465 & 
 0.88800 & 5.376 & 5.48 \\
$\approx\!0.06$ & 0.0036  / 0.018 & 7.47 & $48^3\times144$ & 751 & 
 0.88788 & 5.353 & 4.49 \\
$\approx\!0.06$ & 0.0025  / 0.018 & 7.465 & $56^3\times144$ & 768 & 
 0.88776 & 5.330 & 4.39 \\
$\approx\!0.06$ & 0.0018  / 0.018 & 7.46 & $64^3\times144$ & 826 & 
 0.88764 & 5.307 & 4.27 \\
\noalign{\vspace{-0.06cm}}
\hline
$\approx\!0.06$ & 0.0036  / 0.0108 & 7.46 & $64^3\times144$ & 601 & 
 0.88765 & 5.307 & 5.96 \\
\noalign{\vspace{-0.06cm}}
\hline
$\approx\!0.045$ & 0.0028  / 0.014 & 7.81 & $64^3\times192$ & 801 & 
 0.89511 & 7.208 & 4.56 \\
\hline
\end{tabular}
\end{center}
\vspace{-0.2cm}
\caption{List of ensembles used in this study, with $u_0$ the
tadpole factor and $r_1/a$ the scale from the heavy quark potential.
The $r_1/a$ values are ``mass-independent,'' obtained
by using an interpolating fit to adjust the sea quark masses to 
their physical values \cite{Bazavov:2009bb}. 
Errors in these $r_1/a$ values are roughly 3 to 10 in the last
digit. Configuration
numbers with asterisks have been updated since last year.}
\label{tab:ensembles}
\vspace{-0.2cm}
\end{table}


As always in MILC staggered simulations, we
take the fourth root of the fermion determinant in order
to eliminate the effect of unwanted degrees of freedom (``tastes'') in 
the lattice generation. 
Recent work (see Ref.~\cite{Bazavov:2009bb} and \cite{GOLTERMAN-ROOTREVIEW} for reviews)
makes a strong case that the procedure does indeed produce the desired
theory in the continuum limit.

At the fairly small lattice spacings considered here, chiral effects
due to taste-violations are relatively small but are nevertheless
not negligible.
We take these effects into account at NLO in our chiral fits
by using rooted staggered \chpt\ (\rschpt) 
\cite{Aubin:2003mg}.
An \rschpt\ calculation of the two-loop chiral logarithms does
not exist, so at NNLO we use the continuum partially quenched chiral logarithms
\cite{Bijnens:2004hk}, with the  the root mean square taste average of
the pion mass as argument. In what we call ``low-mass'' fits
(using the low-$m'_s$ ensembles,  with low valence masses),
neglect of taste-violating
effects is justified at NNLO because such
effects are smaller than the mass effects that we
keep. Thus such NNLO low-mass fits are ``systematic'' in the sense of
\chpt. See Ref.~\cite{SU3:2009} for more details.

In  Ref.~\cite{SU3:2009}, the low-mass fits used prior widths to constrain the NNLO LECs 
to be ${\cal O}(1)$ in natural units \cite{Aubin:2004fs}; LO and NLO LECs were unconstrained.
With the current complete data set, we can dispense with the constraints and still
obtain convergent fits. The data alone now determines
the NNLO LECs. It is therefore possible to quote physical results for these quantities,
although, not surprisingly, the errors at NNLO are quite large.
Eliminating constraints at NNLO also tends to increase the errors of the NLO LECs, because
there can be more trade-off between the two orders in the fit.  Clearly the new errors
are more conservative. Fortunately, eliminating constraints has very little effect on
quantities extrapolated to physical meson masses, such as the quark masses and the decay
constants $f_\pi$ and $f_K$.  Such quantities are determined largely by the lattice
data (which now goes to quite low light quark masses) and are not very sensitive to
how the fits divide up contributions by orders in \chpt.

Another feature of the current analysis also tends to increase, relative
to Ref.~\cite{SU3:2009}, our estimate of errors
of certain quantities, in particular, the values of quantities in the 3-flavor chiral limit.
Define the chiral coupling relevant at two-loops as $1/(16\pi^2f_{NNLO}^2)$. The value
of the decay constant $f_{NNLO}$ in the coupling 
is ambiguous {\it a priori}\/ 
because differences in $f_{NNLO}$ (between,
say, the 3-flavor chiral-limit value, $f_3$, and the physical $f_\pi$) are higher order.
In Refs.~\cite{Aubin:2004fs,Bazavov:2009bb}, which did not include NNLO
chiral logs, the corresponding issue already arose at NLO. We found there that
using the ``bare'' coupling with $f_{NLO}=f_3$ did not give acceptable fits,
and it was necessary to use a  ``physical'' NLO coupling (with $f_\pi \ltwid f_{NLO} \ltwid f_K$)
to get good fits to our lattice data.  Similarly, here we cannot obtain good fits 
of our low-mass data with  $f_{NNLO}=f_3$:
The best such fits have confidence level ${\rm CL} < 0.03$, and
in addition require unreasonably large lattice-spacing dependence. On the other hand,
using $f_{NNLO}\approx f_\pi$ gives good fits, and indeed a value close to $f_\pi$ is chosen
by the data if $f_{NNLO}$ in the NNLO terms is allowed to be a free parameter.

\begin{figure}
\begin{center}
\begin{tabular}{c c}
\hspace{-0.5truecm}
\includegraphics[width=0.48\textwidth]{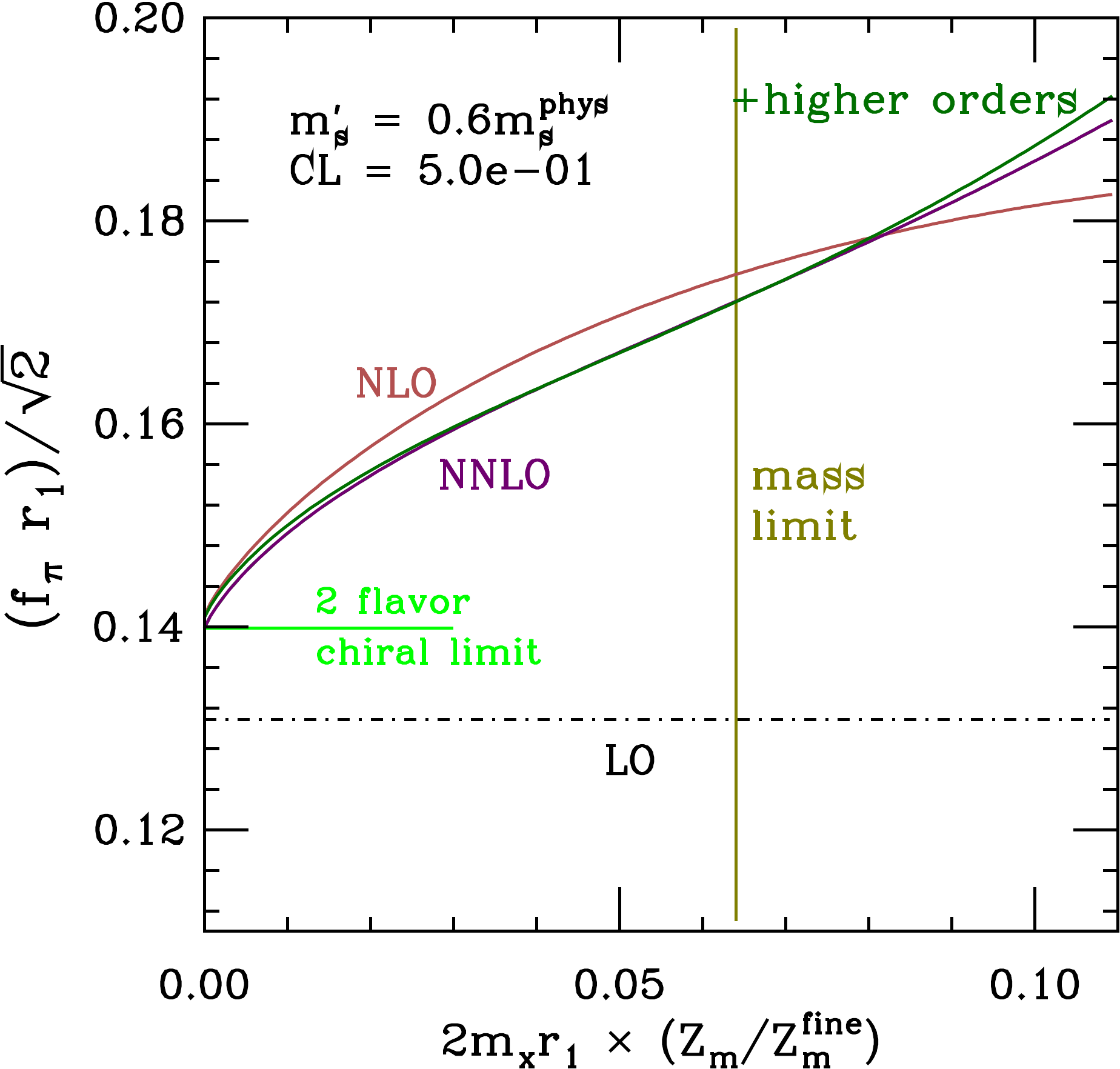}
&
\includegraphics[width=0.465\textwidth]{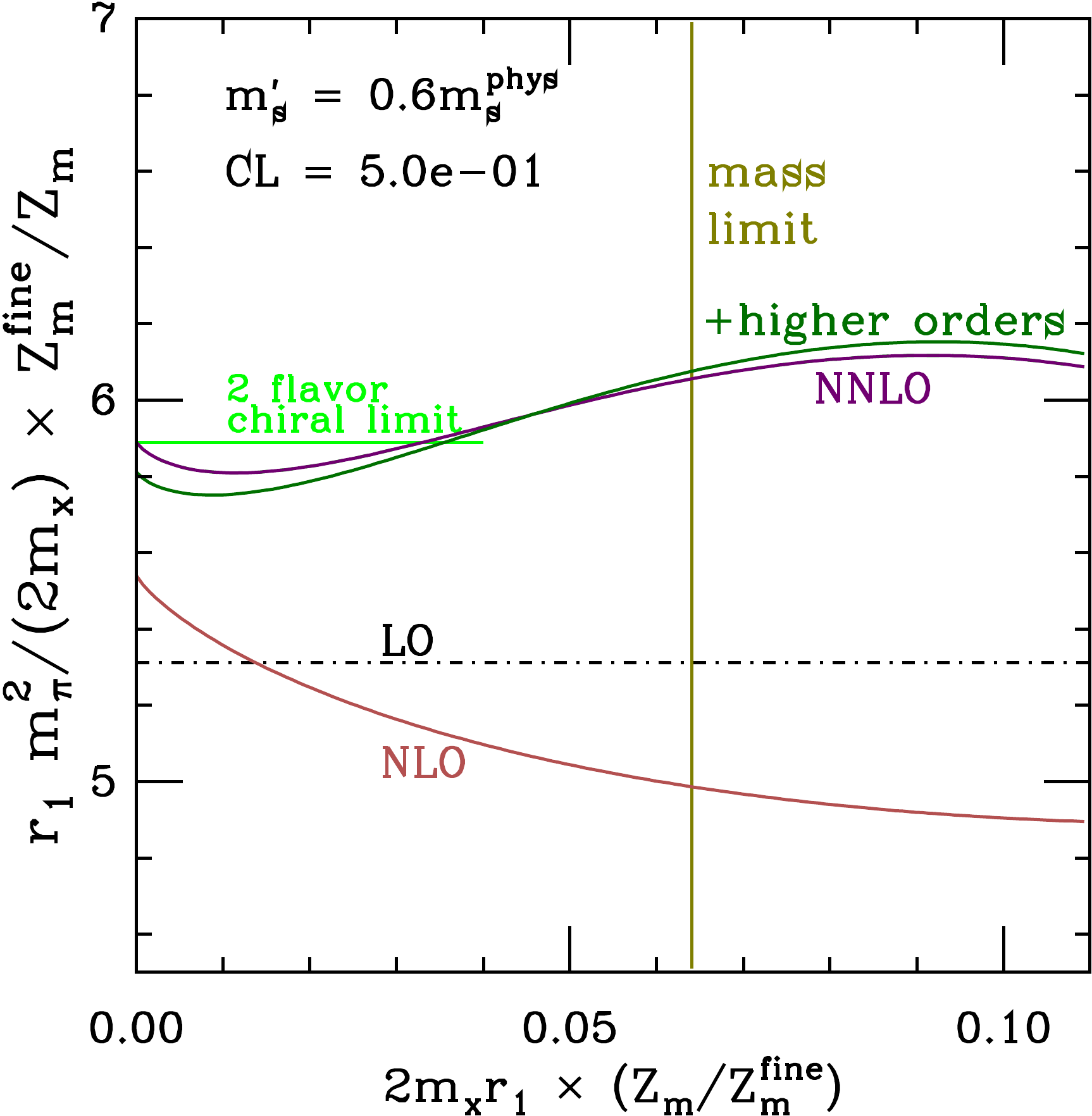}
\end{tabular}
\end{center}
\vspace{-0.7truecm}
\caption{Convergence of ``type-A'' NNLO low-mass SU(3) chiral fits for $f_\pi$ (left)
and $m_\pi^2/2m_x$ (right) as a function of the (degenerate) valence quark mass
$m_x$. The strange sea quark mass $m_s'$ is held fixed at $\approx 0.6 m_s$. The lines
labeled  LO, NLO, and NNLO show the \chpt\ contribution up to and including
the indicated order. The line labeled ``+higher orders'' shows the effect
of adding still higher order analytic terms. 
The vertical line labeled ``mass limit'' corresponds to the largest
value of the valence mass (in the units of the abscissa) used in the fits, and is
given by $2 m_x \approx m_s'$.  The fit has been extrapolated to the continuum.
\label{fig:conv_low_mass}}
\vspace{-0.4truecm}
\end{figure}

Fixing the value of $f_{NNLO}$ is done in two different ways in our fits.
Our standard approach (``type-A fits''), used also in Ref.~\cite{SU3:2009}, is to
put, for each ensemble, $f_{NNLO}=\tau f_3^{\rm fit}$, where
$f_3^{\rm fit}$ is the decay constant in the chiral limit on that ensemble, and
$\tau$ is a fixed number. The value of $\tau$ is chosen (iteratively) 
so that $f_{NNLO}$ has the desired value in the continuum limit. For the
central value fits we ensure $f\cong f_\pi$ in the continuum, and other fits
vary $f_{NNLO}$ over a range (roughly
$0.95 f_\pi \ltwid f_{NNLO} \ltwid 1.15 f_\pi$) that gives acceptable CLs.
An alternative approach  (``type-B fits'') that we have tried recently
is simply to set, on all ensembles,
$f_{NNLO}=\rho f_\pi$, where
$f_\pi$ is taken from experiment,
and $\rho$ is a fixed number, again chosen over the range that gives acceptable fits.
The type-B approach completely decouples the value of $f_{NNLO}$ 
from the chiral limit quantity  $f_3^{\rm fit}$ that appears at lower order and should
describe the decay constant in the low-mass regime.

Chiral fitting is then done in two stages.
Systematic low-mass fits are performed first and are  used to determine low
energy constants (LECs) through NNLO: LO parameters $f_3$ and $B_3$ (often called
$F_0\sqrt{2}$ and $B_0$ in the literature),
NLO Gasser-Leutwyler \cite{Gasser:1984gg} parameters $L_i$, and NNLO parameters 
\cite{Bijnens:1999sh,Bijnens:2004hk}
$K_i$ (partially quenched) or $C_i$ (unquenched SU(3)).
The fits include all partially quenched data for pion and
kaon (with lighter than physical strange quark mass) decay constants
and masses (108 points) and have 31 parameters.  The full covariance matrix is
used, and the CL is good (0.50).
 The convergence of \chpt\ in these
fits (type-A) is shown in Fig.~\ref{fig:conv_low_mass}. Although the NLO terms
for $m_\pi^2/2m_x$ seem to be anomalously small, the overall convergence is
satisfactory.

\begin{figure}
\begin{center}
\begin{tabular}{c c}
\hspace{-0.5truecm}
\includegraphics[width=0.50\textwidth]{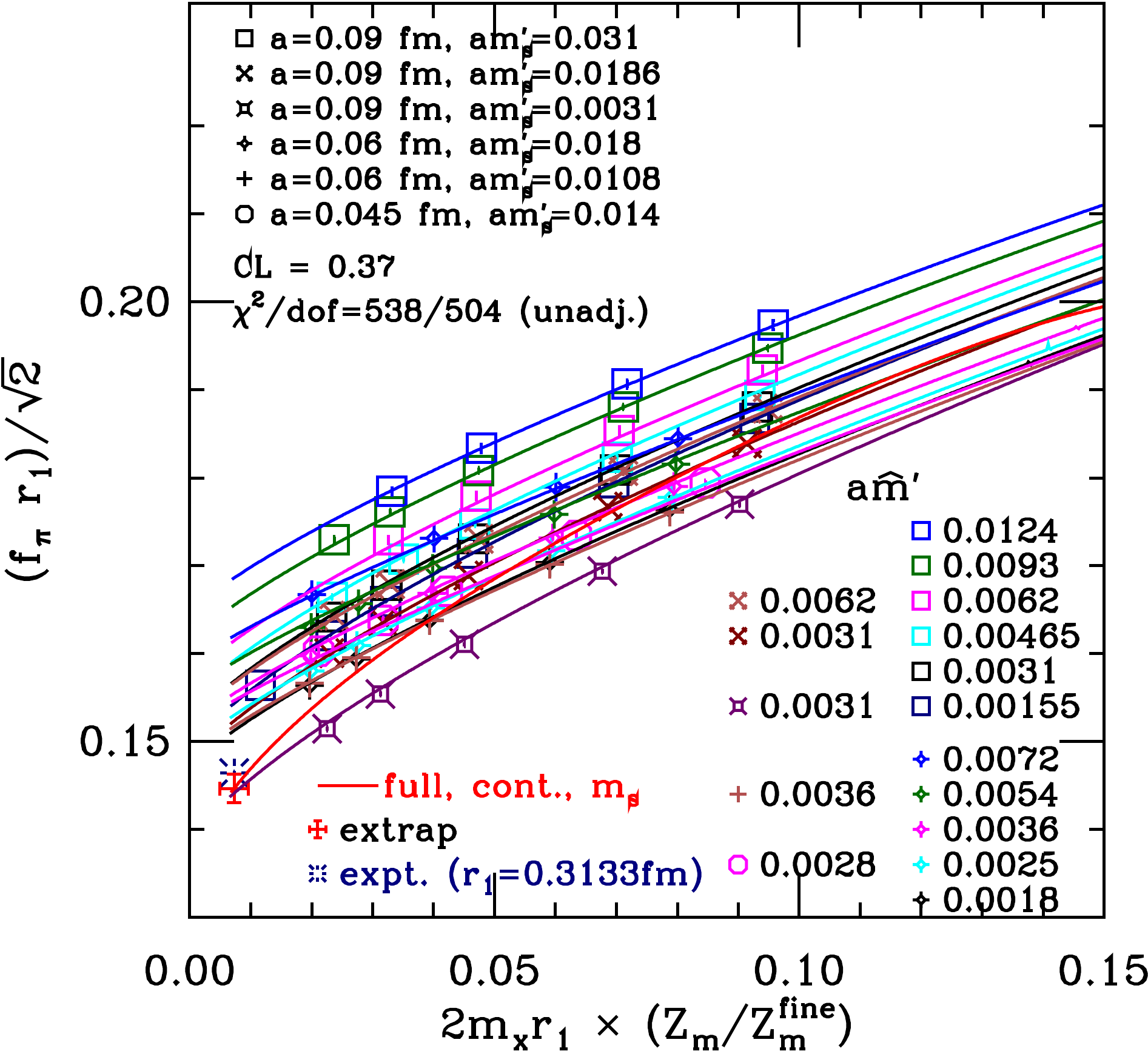}
&
\includegraphics[width=0.47\textwidth]{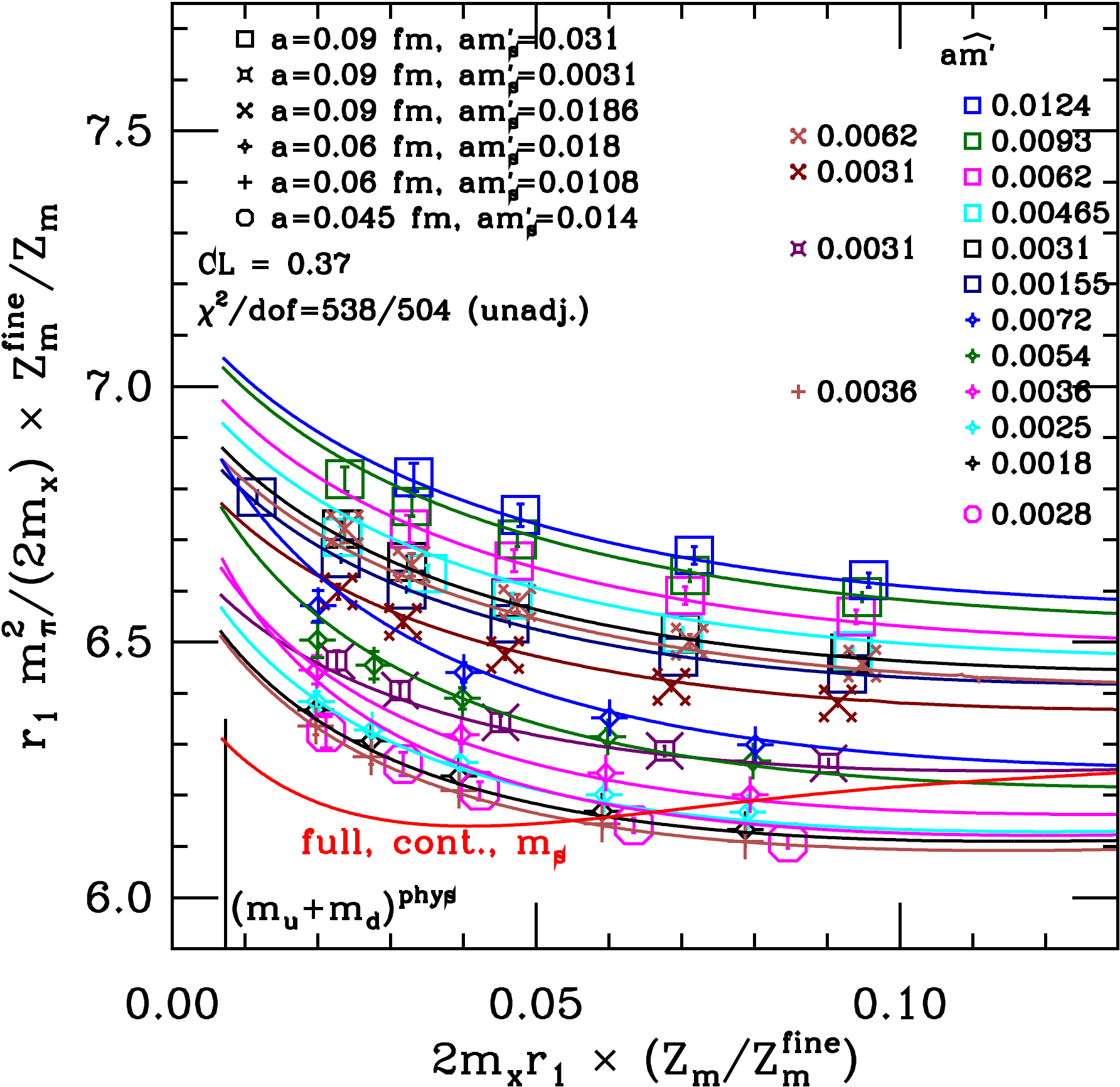} 
\end{tabular}
\end{center}
\vspace{-0.6truecm}
\caption{High-mass, type-A, SU(3) chiral fits, using $r_1=0.3133$. We show partially
quenched data points with degenerate valence masses $m_x$
for the decay constant
(left) and for $m_\pi^2/(2m_x)$ (right). The red lines show the results after extrapolation
to the continuum, putting $m'_s=m_s$, and setting light sea and valence masses equal.  
The decay constant is also extrapolated to the physical mass point and compared with
experiment (blue burst).
\label{fig:high_mass}}
\end{figure}
In the second stage of fitting, the ``high-mass'' fits, all ensembles
listed in Table~\ref{tab:ensembles} are included with the valence
masses restricted to $m_x + m_y \le 1.2 m_s$. 
In the type-A case, the LO, NLO, NNLO
LECs are fixed (for central values) or fixed within statistical errors
at the values from the low-mass fits. N${}^3$LO and N${}^4$LO
analytic terms are included, but not the corresponding logs. These
terms are needed to obtain good CLs, and they allow us to interpolate 
around the (physical) strange
quark mass.  The fact that they are required indicates that SU(3) \chpt\ is not converging
well at these mass values, 
unlike the situation in the low-mass case (see below for more evidence).  However,
since the LO, NLO, and NNLO LECs dominate the chiral extrapolation to the
physical point, the results for decay constants and masses are
rather insensitive to the form of these N${}^3$LO and N${}^4$LO interpolating terms, as long as the
fits are good. 

In the type-B case,  high-mass fits with the LO, NLO, and NNLO
LECs fixed at their low-mass values do not give good CLs.  We can obtain acceptable fits
by allowing these LECs to vary, with prior widths set by the low-mass statistical errors.  But
in that case, several of the LECs move by more than $2\sigma$.  For these reasons the type-B
fits are currently disfavored, although we do include them as alternative chiral fits in estimating
systematic errors.

The (central value, type-A) high-mass fits are illustrated for pions (degenerate valence
points)
in Fig.~\ref{fig:high_mass}, where the scale has been set by $r_1=0.3133(23)\;$fm
\cite{Davies:2009tsa}.
The extrapolations to the continuum and to the physical masses are also shown.  
The agreement of our result, $f_\pi=129.2\pm0.4\pm1.4\;\MeV$, with the experimental value
($130.4(2)\;\MeV$ \cite{PDG}) is good. From now on we set the scale using $f_\pi$,
which gives $r_1 = 0.3106(8)(14)(4)\; {\rm fm}$, where the last error is experimental.

\begin{figure}
\begin{center}
\begin{tabular}{c c}
\hspace{-0.5truecm}
\includegraphics[width=0.50\textwidth]{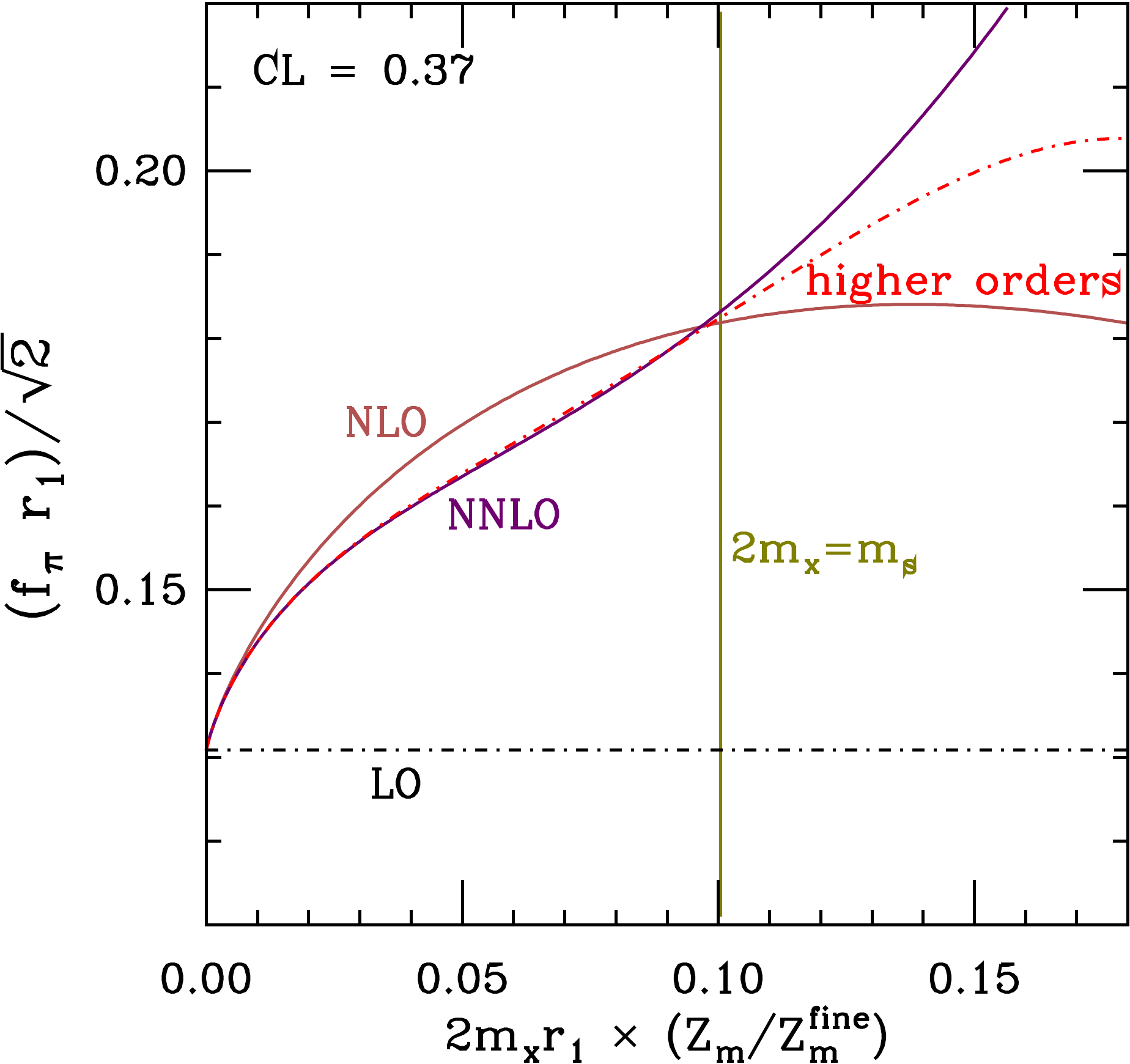}
&
\includegraphics[width=0.50\textwidth]{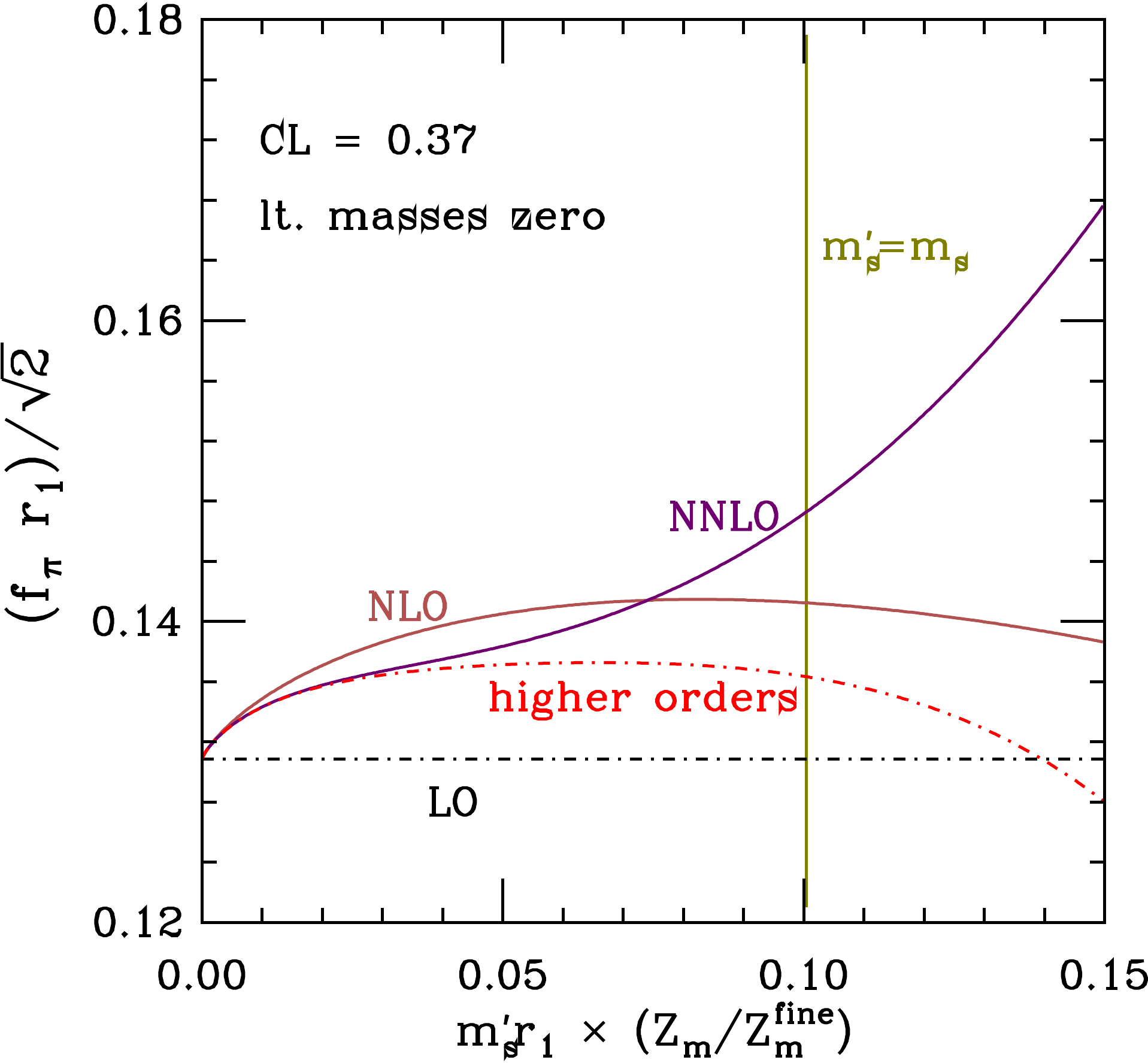}
\end{tabular}
\end{center}
\vspace{-0.6truecm}
\caption{
Convergence of \chpt\ for the decay constant, with the same fit as in
Fig.~2. Left, the degenerate
case ($m'_s ={\hat{m}}' = m_x$) as a function of $m_x$. Right, 2-flavor chiral limit $m_x=\hat m'=0$ as a function of $m'_s$. These fits have been extrapolated to the continuum.
\label{fig:convergence}}
\end{figure}

Figure~\ref{fig:convergence}
shows the convergence of \chpt\ for the 
pion decay constant.  The left plot is for the degenerate 3-flavor case.  Note that 
the convergence is still good significantly beyond the point where $2m_x=m_s$, 
where the meson mass is already close to the physical value of $m_K$. The right plot 
gives the decay constant as a function of the strange
sea quark mass $m'_s$, with the valence and light sea masses extrapolated to zero (the 2-flavor
chiral limit).  The vertical line indicates $m_s'=m_s$, when the kaon is near physical.   Here
the convergence is reasonable only up to about $m_s' =0.6m_s$; by  $m_s' =0.8m_s$ the NNLO
term has the wrong sign, and moves the NLO result {\it away}\/ from the ``higher orders'' line
(which must be close to what the lattice data demands, since the fit is good). Nevertheless,
the NLO contribution remains reasonable; the problem is mainly at NNLO.  This kind of
behavior is not unexpected for an asymptotic expansion. 

For the $m_\pi^2/(2m_x)$ case (not shown), the
convergence of \chpt\ is similar to that of the decay constant.  Again, convergence
in the degenerate 3-flavor case is 
good up to meson masses about 10\% or 15\% higher than the kaon.
In the 2-flavor chiral limit, the behavior 
as a function of $m_s'$ 
appears to be comparable to that of  the decay constant, breaking down before $m_s'=m_s$.

\section{Preliminary results}

The following is a sample of our current results:
\begin{eqnarray}\label{eq:res_fpi_scal}
f_K = 156.1 \pm 0.4\; {}^{+0.6}_{-0.9} \; \MeV \ , &\quad\qquad& 
f_K/f_\pi = 1.197(2)({}^{+3}_{-7}) \ , \nonumber \\
V_{us} = 0.2247({}^{+14}_{-\phantom{0}9})\ ,  &\quad\qquad& 
f_3 = 113.6 \pm 3.6 \pm 7.7 \; \MeV \ ,  \nonumber   \\
f_2 = 123.0 \pm 0.5 \pm 0.7 \; \MeV \ , &\quad\qquad&
2L_8 - L_5 = -0.51(11)({}^{+45}_{-19}) \ , \nonumber \\
2L_6 - L_4 = 0.09(24)({}^{+32}_{-27}) \ , &\quad\qquad& 
L_5 = 1.79(16)({}^{+28}_{-41}) \ , \nonumber  \\
L_4 = 0.19(22)({}^{+57}_{-33}) \ , &\quad\qquad& 
2C_{21}-C_{19}= 1.5(6)({}^{+6}_{-4}) \ ,  \nonumber \\
K_{19} = 3.5\pm1.2\;{}^{+2.3}_{-0.7} \ , &\quad\qquad&
K_{39}-K_{17}= 3.4\pm1.5\;{}^{+1.8}_{-1.6} \ ,  \nonumber 
\end{eqnarray}
\begin{eqnarray}
\bar l_3 = 3.18(50)(89) \ ,  &\quad\qquad& 
\bar l_4 = 4.29(21)(82) \ , \nonumber 
\end{eqnarray}
where errors are statistical and systematic. $f_2$ is the decay constant in the
2-flavor chiral limit.
The NLO LECs $L_i$ are in units of $10^{-3}$, 
and the NNLO LECs $K_i$ and $C_i$ are in units of  $10^{-6}$; both are at
chiral scale $m_\eta$.  Other  $K_i$ and $C_i$ are also of this order of magnitude, but
most have statistical or systematic errors that are more than 100\%.
The scale invariant SU(2) LECs $\bar\ell_{3,4}$ are obtained from the
SU(3) LECs using the
two-loop conversion formulae \cite{Gasser:2007sg}.
There is good agreement between the SU(3) chiral fit results
described here and the results of the SU(2) chiral fits 
\cite{Xining:lat09,Xining:lat10} for all quantities that can be directly compared.

As discussed above, errors for $f_3$ and
the NLO LECs are larger than previous ones \cite{SU3:2009},
while those for $f_K$, $f_K/f_\pi$, $f_2$, as well as for quark masses (not given here),
are similar to or somewhat smaller than before. In general, we use type-A fits
with $f_{NNLO}\cong f_\pi$ for central
values, and alternative chiral fits (including type-B) for systematic error estimates.
The exception is $f_3$, where we have averaged type-A and type-B results
and symmetrized the errors.  This is because the distribution for $f_3$
from type-A and type-B fits is bimodal, with the former giving values toward the high end of the
range, and the latter giving values toward the low end. The large uncertainty in the
3-flavor chiral limit, compared to the physical point, comes ultimately from the fact
that we only have one usable ensemble with three degenerate light flavors, while there
are many with $m'_s\approx m_s$.

We thank J.\ Bijnens for his program to compute the
partially quenched NNLO chiral logs.

\end{document}